\documentclass{llncs}

\newcommand{\ieeeonly}[1]{}
\newcommand{\lncsonly}[1]{#1}
\newcommand{\articleonly}[1]{}
\newcommand{\acmonly}[1]{}
\newcommand{\svonly}[1]{}

\newcommand{\mycomment}[1]{}

\newcommand{\fullonly}[1]{#1} 
\newcommand{\shortonly}[1]{} 

\lncsonly{\pagestyle{headings}}

\acmonly{
\settopmatter{printacmref=true}
\fancyhead{}

\def\BibTeX{{\rm B\kern-.05em{\sc i\kern-.025em b}\kern-.08emT\kern-.1667em\lower.7ex\hbox{E}\kern-.125emX}}
    
}

\usepackage{hyperref}
\usepackage{graphicx}
\usepackage{xspace}
\usepackage{amsmath}  
\usepackage[subtle,mathspacing=normal]{savetrees} 

\usepackage{multirow}
\usepackage{graphicx}

\articleonly{
\usepackage{setspace}
\setstretch{1.1}
\usepackage{fullpage} 
}

\articleonly{}
\lncsonly{}
\acmonly{}

\newcommand{\tuple}[1]{\langle #1\rangle}
\newcommand{\mean}[1]{\left[ \! \left[ #1 \right]\! \right]}

\newcommand{\intersect}{\cap}

\newcommand{\cm}{{\it CM}}
\newcommand{\om}{{\it OM}}
\newcommand{\type}{{\rm type}}
\newcommand{\scond}{{\rm sCond}}
\newcommand{\rcond}{{\rm rCond}}
\newcommand{\con}{{\rm con}}
\newcommand{\acts}{{\rm acts}}
\newcommand{\Act}{{\it Act}}
\newcommand{\nav}{{\rm nav}}
\newcommand{\eval}{{\rm tval}}
\newcommand{\val}{{\it val}}

\newcommand{\op}{{\it op}}
\newcommand{\Rules}{{\it Rules}}
\newcommand{\spa}{{\it AU}}

\newcommand{\ind}{\hspace*{0.7em}}

\newcommand{\paths}{{\rm paths}}

\newcommand{\getpath}{{\rm path}}
\newcommand{\getsign}{{\rm sign}}
\newcommand{\getval}{{\rm val}}

\newcommand{\ifstmt}{{\bf if}}
\newcommand{\thenstmt}{{\bf then}}
\newcommand{\elsestmt}{{\bf else}}

\newcommand{\forloop}{{\bf for}}
\newcommand{\foreachloop}{{\bf for each}}
\newcommand{\whileloop}{{\bf while}}

\newcommand{\breakstmt}{{\bf break}}

\newcommand{\forloopin}{{\bf in}}

\newcommand{\com}[1]{{\it #1}}

\newcommand{\union}{\cup}

\newcommand{\synSim}{{\rm syn}}

\newcommand{\acSim}{\synSim_{\rm ac}}

\newcommand{\Qpol}{Q_{\rm pol}}

\newcommand{\average}{{\rm average}}

\newcommand{\wsc}{{\rm WSC}}

\begin{document}

\acmonly{
\copyrightyear{2020} 
\acmYear{2020} 
\setcopyright{acmcopyright}
\acmConference[SACMAT '20]{25th ACM Symposium on Access Control Models and Technologies}{June 10--12, 2020}{Barcelona, Spain}
\acmBooktitle{25th ACM Symposium on Access Control Models and Technologies (SACMAT '20), June 10--12, 2020, Barcelona, Spain}
\acmPrice{15.00}
\acmDOI{10.1145/3381991.3395619}
\acmISBN{978-1-4503-7568-9/20/06}
\fancyhead{}
}

\newcommand{\thanksText}{This material is based on work supported in part by NSF grant
    CCF-1954837 
    and ONR grant N00014-20-1-2751. 
     }
    
\title{
Learning Attribute-Based and Relationship-Based Access Control Policies with Unknown Values\ieeeonly{\thanks{\thanksText}}\lncsonly{\thanks{\thanksText}}
}

\ieeeonly{\author{
\IEEEauthorblockN{Thang Bui and Scott D. Stoller}
\IEEEauthorblockA{Department of Computer Science, Stony Brook University, USA}}}

\articleonly{\author{Thang Bui and Scott~D.~Stoller\\
  Department of Computer Science, Stony Brook University, USA}}

\lncsonly{\author{Thang Bui \and Scott~D.~Stoller}
  \institute{Department of Computer Science, Stony Brook University, USA}}

\acmonly{
\author{Thang Bui}
\affiliation{
\institution{Stony Brook University}}
\email{thang.bui@stonybrook.edu}
\author{Scott~D.~Stoller}
\affiliation{
\institution{Stony Brook University}}
\email{stoller@cs.stonybrook.edu}
}


\svonly{\author{Thang Bui \and Scott~D.~Stoller}
\institute{T. Bui \and S. D. Stoller \at Stony Brook University, USA\\ \email{stoller@cs.stonybrook.edu}}
\date{Received: date / Accepted: date}}

\newcommand{\abstracttext}{ %
Attribute-Based Access Control (ABAC) and Relationship-based access control (ReBAC) provide a high level of expressiveness and flexibility that promote security and information sharing, by allowing policies to be expressed in terms of attributes of and chains of relationships between entities.  Algorithms for learning ABAC and ReBAC policies from legacy access control information have the potential to significantly reduce the cost of migration to ABAC or ReBAC.

This paper presents the first algorithms for mining ABAC and ReBAC policies from access control lists (ACLs) and incomplete information about entities, where the values of some attributes of some entities are unknown.  We show that the core of this problem can be viewed as learning a concise three-valued logic formula from a set of labeled feature vectors containing unknowns, and we give the first algorithm (to the best of our knowledge) for that problem.
}

\acmonly{
\begin{abstract}
\abstracttext
\end{abstract}}


\acmonly{}


\maketitle
\ieeeonly{
\begin{abstract}
\abstracttext
\end{abstract}}
\lncsonly{
\begin{abstract}
\abstracttext
\end{abstract}}
\articleonly{
\begin{abstract}
\abstracttext
\end{abstract}}
\svonly{
\begin{abstract}
\abstracttext
\end{abstract}}


\section{Introduction}
\label{sec:intro}

Relationship-based access control (ReBAC) extends the well-known attribute-based access control (ABAC) framework by allowing access control policies to be expressed in terms of chains of relationships between entities, as well as attributes of entities.  This significantly increases the expressiveness and often allows supporting more natural policies.  High-level access control policy models such as ABAC and ReBAC are becoming increasingly widely adopted, as security policies become more dynamic and more complex, and because they promise long-term cost savings through reduced management effort. ABAC is already supported by many enterprise software products.  Forms of ReBAC are supported in popular online social network systems and are being studied and adapted for use in more general software systems as well.

The up-front cost of developing an ABAC or ReBAC policy can be a significant barrier to adoption.  {\em Policy mining} (a.k.a. policy learning) algorithms have the potential to greatly reduce this cost, by automatically producing a draft high-level policy from existing lower-level data, such as access control lists or access logs.
There is a substantial amount of research on role mining\fullonly{ \cite{mitra2016survey,das2018}} and 
a small but growing literature on ABAC policy mining\fullonly{ \cite{xu15miningABAC,xu14miningABAClogs,medvet2015,mocanu2015,das2018,sparselogs2018,iyer2018,karimi2018,cotrini2019,law20fastlas}}\shortonly{, surveyed in \cite{das2018}}, and ReBAC policy mining \cite{bui17mining,bui18mining,bui19mining,bui19sacmat,iyer2019,bui20decision,iyer2020active}.

The basic ABAC (or ReBAC) policy mining problem is: Given information about the attributes of entities in the system, and the set of currently granted permissions; Find an ABAC (or ReBAC) policy that grants the same permissions using concise, high-level ReBAC rules.  Several papers consider a variant of this problem where the information about permissions is incomplete \cite{xu14miningABAClogs,bui18mining,sparselogs2018,iyer2020active,law20fastlas}.  However, all existing works on ABAC and ReBAC policy mining assume that the attribute (and relationship) information is complete, i.e., all attributes of all entities have known values.  Unfortunately, in most real-world data, some attribute values are unknown (a.k.a. missing).  Bui et al. \cite{bui17mining,bui18mining,bui19mining,bui19sacmat} allow an attribute to have the special value ``bottom'', which is analogous to {\tt None} in Python.  It is different from unknown.  For example, for a field Student.advisor with type Faculty, bottom (or {\tt None}) means the student lacks an advisor, while unknown means we don't know whether the student has an advisor or, if they have one, who it is.  Xu and Stoller \cite{xu15miningABAC} consider ABAC mining from noisy attribute data, where some of the given attribute values are incorrect; this is also different, because the input does not specify which ones are incorrect.

{\em This paper proposes the first algorithms for mining ABAC or ReBAC policies when some attribute values are unknown.}  We present our algorithm in the context of ReBAC mining because ReBAC is more general than  ABAC.  Our algorithm can easily be restricted to mine ABAC policies instead, simply by limiting the length of path expressions that it considers.

Our main algorithm, called DTRMU$^-$ (Decision-Tree ReBAC Miner with Unknown values and negation), produces policies in ORAL2$^-$, an object-oriented ReBAC language introduced by Bui and Stoller \cite{bui20decision}.  We chose ORAL2$^-$ because it is more expressive than other policy languages that have been used in work on ReBAC mining.  In ORAL2$^-$, relationships are expressed using object attributes (fields) that refer to other objects, and chains of relationships between objects are described by {\em path expressions}, which are sequences of attribute dereferences.  A policy is a set of rules.  A rule is essentially a conjunction of conditions on the {\em subject} (an object representing the issuer of the access request), conditions on the {\em resource} (an object representing the resource to be accessed), and constraints relating the subject and resource; the subject may perform a specified action on the resource if the conditions and constraints are satisfied.  An example of a condition is subject.employer = LargeBank; an example of a constraint is subject.department $\in$ resource.project.departments.  ORAL2$^-$ also supports negation, so conditions and constraints can be negated, e.g.,  subject.employer $\ne$ LargeBank.  We also give an algorithm, called DTRMU, that mines policies in ORAL2, which is the same as ORAL2$^-$ except without negation.  Deciding whether to include negation in the policy language involves a trade-off between safety and conciseness, as discussed in \cite{bui20decision}; different organizations might make different decisions, and we support both.
 
A policy can be viewed, roughly speaking, as a logical formula in disjunctive normal form (DNF), namely, the disjunction of the conjunctions (of conditions and constraints) in the rules.  Bui and Stoller \cite{bui20decision} exploited this view to reduce the core of the ReBAC policy mining problem to decision-tree learning; note that a decision tree compactly represents a logical formula in DNF, where each conjunction contains the conditions labeling the nodes on a path from the root to a leaf labeled ``true'' (corresponding to ``permit'').

Our algorithms are built on the insight that the core of the ReBAC policy mining problem in the presence of unknown attribute values can be reduced to the general problem of {\em learning a formula in Kleene's three-valued logic} \cite{kleene1950,3vlogic}, rather than traditional Boolean logic.  Three-valued logic allows three truth values: true ($T$), false ($F$), and unknown ($U$).  With three-valued logic, we can assign the truth value $U$ to conditions and constraints involving unknown attribute values.  Could the need for three-valued logic be avoided by regarding them as false instead?  No, because if we stick with Boolean logic, and declare that (say) the condition subject.employer = LargeBank is false when the employer is unknown, then we are forced to conclude that its negation, $\neg({\rm subject.employer = LargeBank})$, is true when the employer is unknown, and this is clearly unsafe.  Note that SQL uses three-valued logic to deal with {\tt null} (i.e., missing) values for similar reasons.


Surprisingly, we could not find an existing algorithm for learning a concise three-valued logic formula from a set of labeled feature vectors containing unknowns.  Therefore, we developed an algorithm to solve this general problem, based on learning multi-way decision trees, and then adapted Bui and Stoller's Decision-Tree ReBAC Mining algorithms (DTRM and DTRM$^-$) to use that algorithm.  We adopted their decision-tree based approach, because their algorithms are significantly faster, achieve comparable policy quality, and can mine policies in a richer language than other ReBAC mining algorithms such as FS-SEA* \cite{bui19sacmat} and Iyer et al.'s algorithm \cite{iyer2019}, as demonstrated by their experiments \cite{bui20decision}.



We performed two series of experiments on several ReBAC policies.  The first series of experiments compares our algorithms with Bui and Stoller's DTRM and DTRM$^-$ algorithms, and shows that, {\em on policies where all attribute values are known, our algorithms are equally effective at discovering the desired ReBAC rules, produce policies with the same quality, and have comparable running time}.  The second series of experiments, on policies containing a varying percentage of unknown values, shows that {\em our algorithms are effective at discovering the desired ReBAC rules, even when a significant percentage of attribute values are unknown}.

In summary, the main contributions of this paper are the first ABAC and ReBAC policy mining algorithms that can handle unknown attribute values, and, to the best of our knowledge, the first algorithm for learning a concise three-valued logic formula from a set of labeled feature vectors containing unknowns.  Directions for future work include extending our algorithms to deal with incomplete information about permissions and extending them to ``fill in'' missing attribute values, guided by the permissions.  Another is developing incremental algorithms to efficiently handle policy changes.  Note that, as usual in ABAC and ReBAC policy mining, changes to attribute data (known values changing, or unknown values becoming known) do not require learning a new policy, except in the infrequent case that the current policy does not grant the desired permissions.


\section{Learning Three-Valued Logic Formulas}
\label{sec:general-problem}
\subsection{Problem Definition}
\label{sec:general-problem-def}

We consider the problem of learning a formula in Kleene's three-valued logic from a set of labeled feature vectors.  The feature values and the labels are truth values in three-valued logic, namely, true ($T$), false ($F$), and unknown ($U$).  In this setting, the features would usually be called ``propositions'', and the feature vectors would usually be called ``interpretations'', but we prefer to use more general terminology.  The conjunction, disjunction, and negation operators are extended to handle unknown, in a natural way \cite{kleene1950,3vlogic}.  For example, $T \lor U$ evaluates to $T$, while $T \land U$ evaluates to $U$.

We require that the set of labeled feature vectors is monotonic, in the sense defined below, otherwise there would be no three-valued logic formula that represents it.  For a feature vector $v$ and feature $f$, let $v(f)$ denote the value of feature $f$ in $v$.  For a formula $\phi$, let $\phi(v)$ denote the truth value of $\phi$ for $v$, i.e., the result of evaluating $\phi$ using the truth values in $v$.



For truth values $t_1$ and $t_2$, $t_1 \le t_2$ iff $t_1=t_2$ or $t_1 = U$.  This is sometimes called the {\em information ordering}; it captures the idea that $U$ provides less information than $T$ and $F$.  For feature vectors $v_1$ and $v_2$, $v_1 \le v_2$ iff $v_1(f) \le v_2(f)$ for every feature $f$.  A basic fact of three-valued logic is that every formula, regarded as a function from feature vectors to truth values, is monotonic with respect to the information ordering, i.e., for all feature vectors $v_1$ and $v_2$, if $v_1 \le v_2$ then $\phi(v_1) \le \phi(v_2)$.

A set $S$ of labeled feature vectors is {\em monotonic} iff, for all $(v_1, \ell_1)$ and $(v_2, \ell_2)$ in $S$,
if $v_1 \le v_2$ then $\ell_1 \le \ell_2$.  This ensures $S$ can be represented by a formula.

The \textit{three-valued logic formula learning problem} is: given a monotonic set $S$ of labeled feature vectors, where the feature values and labels are truth values in three-valued logic, find a three-valued logic formula $\phi$ in disjunctive normal form (DNF) that exactly characterizes the feature vectors labeled $T$, i.e., for all $(v, \ell)$  in $S$, $\phi(v) = T$ iff $\ell=T$.

A stricter variant of this problem requires that $\phi$ preserve all three truth values, i.e., for all $(v, \ell)$  in $S$, $\phi(v) = \ell$.  We adopt the looser requirement above, because when a formula ultimately evaluates to unknown, this outcome is conservatively treated the same as false in many application domains including security policies and SQL queries, and adopting the looser requirement allows smaller and simpler formulas.  Note that distinguishing $U$ and $F$ is still critical during evaluation of formulas and their subformulas, for the reasons discussed in Section \ref{sec:intro}.
The stricter variant of the problem would be relevant in a security policy framework, such as XACML, that allows policies to return indeterminate results; this is relevant mainly when composing policies, since an indeterminate result typically still results in a denial at the top level.

\subsection{Learning a Multi-Way Decision Tree}
\label{sec:algorithm:build-tree}

Since we are dealing with three truth values, we need multi-way decision trees, instead of binary trees. Each internal node is labeled with a feature.  Each outgoing edge of an internal node corresponds to a possible value of the feature.  Each leaf node is labeled with a classification label, which in our setting are also truth values. A feature vector is classified by testing the feature in the root node, following the edge corresponding to the value of the feature to reach a subtree, and then repeating this procedure until a leaf node is reached. A sample decision tree is shown in Section \ref{sec:mining-algorithm}.



Our algorithm uses C4.5 \cite{C45}, a well-known decision tree learning algorithm, to build a multi-way decision tree that correctly classifies a given set $S$ of labeled feature vectors.  It builds a decision tree by recursively partitioning feature vectors in the dataset $S$, starting from a root node associated with the entire dataset. It chooses (as described below) a feature to test at the root node, creates a child node for each possible outcome of the test,  partitions the set of feature vectors associated with the root node among the children, based on the outcome of the test, and recursively applies this procedure to each child.  The recursion stops when all of the feature vectors associated with a node have the same classification label or when there is no feature vector associated with a node (the leaf node is labeled with False in this case).  At each node $n$, the algorithm evaluates a scoring criteria for each of the remaining features (i.e., features that have not been used for splitting at an ancestor of $n$) and then chooses the top-ranked feature. C4.5 uses information gain as the scoring criteria.


\subsection{Algorithm for Learning a Three-Valued Logic Formula}
\label{sec:mining-algorithm:extract-formula}

The algorithm is presented as pseudocode in Figure \ref{fig:general-alg}, with explanations inlined in comments.  It iterates to build a formula $D$ in DNF satisfying the requirements.  For convenience, we represent $D$ as a set of conjunctions; the desired formula is the disjunction of the conjunctions in $D$.  For a path $p$ through a decision tree from the root to a leaf, let conj($p$) be a conjunction of conditions on the features associated with internal nodes on that path; specifically, if the path passes through a node labeled with feature $f$ and follows the out-edge labeled $T$, $F$ or $U$, then $f$, $\neg f$, or $f=U$, respectively, is included as a conjunct.  Although the algorithm uses conditions of the form $f=U$ in intermediate conjunctions, they need to be eliminated, because $f=U$ is not a formula in three-valued logic; furthermore, three-valued logic does not contain any formula equivalent to $f=U$, because this condition is not monotonic (in other words, it does not satisfy the monotonicity property of formulas stated above). A formula $\phi$ is {\em valid} with respect to a set $S$ of labeled feature vectors, denoted valid$(\phi,S)$, if it does not mis-evaluate any feature vectors as true, i.e., for every feature vector $v$ in $S$ labeled $F$ or $U$, $\phi(v)$ is $F$ or $U$. A formula $\phi$ {\em covers} $S$ if $\phi(v)$ is $T$ for every feature vector in $S$ labeled $T$.
An example of how the algorithm works appears in Section \ref{sec:mining-algorithm}.


\begin{figure}[tbp]
\begin{tabular}[t]{@{}l@{}}
$S$ = the given set of labeled feature vectors\\
$D = \emptyset$  // \com{the desired formula in DNF, represented as a set of conjunctions}\\
$B = \emptyset$ // \com{set of black-listed features}\\
$iter$ = 0      // \com{number of iterations of tree learning}\\
\whileloop\ $D$ does not cover $S$ and $iter < max\_iter$\\ 
\ind // \com{add disjuncts until $D$ covers $S$ or max\_iter is reached}\\
\ind $S'$ = $S\setminus \{ (v,T) \;|\; D(v)=T \}$  // \com{remove feature vectors covered by $D$}\\
\ind Use C4.5 to learn a multi-way decision tree $dt$ for $S'$, without using features in $B$\\
\ind $D'$ = set containing conj($p$) for each path $p$ through $dt$ from the root to a leaf labeled $T$\\
\ind // \com{eliminate conjuncts of the form $f=U$}\\
\ind\foreachloop\ conjunction $c$ in $D'$ that contains a condition of the form $f=U$\\
\ind\ind $c' = c$; remove $c$ from $D'$\\
\ind\ind\foreachloop\ condition $f_u$ of the form $f=U$ \forloopin\ $c'$\\
\ind\ind\ind $c''=$ formula obtained from $c'$ by removing $f_u$\\
\ind\ind\ind\ifstmt\ valid($c'', S$) \thenstmt\ $c' = c''$ // \com{successfully removed $f_u$}  \\ 
\ind\ind\ind\elsestmt\\
\ind\ind\ind\ind // \com{$f_u$ cannot simply be removed; try to replace it with another condition}\\
\ind\ind\ind\ind $F_r$ = set containing features not used in $c$, and the negations of those features\\   
\ind\ind\ind\ind\foreachloop\ $f_1$ \forloopin\ $F_r$\\
\ind\ind\ind\ind\ind $c''=$ formula obtained from $c'$ by replacing $f_u$ with $f_1$ \\
\ind\ind\ind\ind\ind\ifstmt\ valid($c'', S$) $\land$ ($D'\cup\{c''\}$ covers $S'$)\\
\ind\ind\ind\ind\ind\ind $c' = c''$ // \com{successfully replaced $f_u$ with $f_1$}\\
\ind\ind\ind\ind\ind\ind\breakstmt\\
\ind\ind\ifstmt\ $c'$ does not contain any conditions of the form $f=U$ \thenstmt\ add $c'$ to $D'$\\
\ind\ind\elsestmt\\
\ind\ind\ind // \com{some $f=U$ conditions in $c$ couldn't be eliminated or replaced.}\\
\ind\ind\ind // \com{discard $c$, and blacklist features used in its $f=U$ conditions.}\\
\ind\ind\ind\foreachloop\ condition of the form $f=U$ \forloopin\ $c$\\
\ind\ind\ind\ind add $f$ to $B$\\
\ind $D = D \union D'$\\
\ind $iter = iter + 1$\\ 
\ifstmt\ $D$ does not cover $S$\\
\ind // \com{max iterations was exceeded. cover the remaining feature vectors one at a time.}\\
\ind $uncov$ =  $\{ v \;|\; (v,T) \in S \land D(v)\ne T \}$ // \com{uncovered feature vectors}\\
\ind\foreachloop\ feature vector $v$ \forloopin\  $uncov$\\
\ind\ind $c$ = conjunction containing the conjunct $f$ for each feature $f$ s.t. $v(f)=T$\\
\ind\ind\ind and the conjunct $\neg f$ for each feature $f$ s.t. $v(f)=F$\\
\ind\ind // \com{note that $c(v)=T$, and monotonicity of $S$ ensures valid$(c,S)$ holds}\\
\ind\ind add $c$ to $D$\\
// \com{remove redundant disjuncts from $D$}\\
\foreachloop\ conjunction $c$ \forloopin\ $D$\\
\ind\ifstmt\ the set of conjuncts in $c$ is a superset of the set of conjuncts in another element of $D$\\
\ind\ind remove $c$ from $D$
\end{tabular}
\caption{Algorithm for learning a three-valued logic formula.}
\label{fig:general-alg}\vspace{-1.5ex}
\end{figure}

\section{Policy Language with Unknown Attribute Values}
\label{sec:language}

We adopt Bui et al.’s ORAL2$^-$ \cite{bui20decision} ReBAC policy language and modify it to handle unknown attribute values.  It contains common ABAC constructs, similar to those in \cite{xu15miningABAC}, plus path expressions.  ORAL2$^-$ can easily be restricted to express ABAC policies by limiting the maximum length of path expressions to 1. We give a brief overview of the language (for details, see \cite{bui20decision}) and focus on describing the changes to handle unknown values.  The largest changes are to the definitions of path dereferencing (see the definition of $\nav$) and the definitions of truth values of conditions and constraints.

A {\em ReBAC policy} is a tuple $\pi=\tuple{\cm, \om, \Act, \Rules}$, where $\cm$ is a class model, $\om$ is an object model, $\Act$ is a set of actions, and $\Rules$ is a set of rules.

A {\em class model} is a set of class declarations. 
Each field has a {\em type}, which is a class name or ``Boolean'', and a {\em multiplicity}, which specifies how many values may be stored in the field and is ``one'' (also denoted ``1''), ``optional'' (also denoted ``?''), or ``many'' (also denoted ``*'', meaning any number).  Boolean fields always have multiplicity 1.  Every class implicitly contains a field ``id'' with type String and multiplicity 1. 

An {\em object model} is a set of objects whose types are consistent with the class model and with unique values in the id fields. Let $\type(o)$ denote the type of object $o$. 
The value of a field with multiplicity ``many'' is a set of values.
The value of a field with multiplicity ``one'' or ``optional'' is a single value.
The value of a field with multiplicity ``optional'' is a value of the specified type or {\tt None} (called ``bottom'' in \cite{bui20decision}). The value of any field can also be the special value {\tt unknown}, indicating that the actual value is unknown (missing).  The difference between {\tt None} and {\tt unknown} is explained in Section \ref{sec:intro}.  {\tt unknown} cannot appear in a set of values in the object model, but it may appear in sets of values constructed by our algorithm.  Note that we distinguish {\tt unknown} (a placeholder used in object models) from $U$ (a truth value in three-valued logic).

A {\em path} is a sequence of field names, written with ``.'' as a separator.  
A {\em condition} is a set, interpreted as a conjunction, of atomic conditions or their negations.
An {\em atomic condition} is a tuple $\tuple{p, \op, \val}$, where $p$ is a non-empty path, $\op$ is an operator, either ``in'' or ``contains'', and $\val$ is a constant value, either an atomic value (if $\op$ is ``contains'') or a set of atomic values (if $\op$ is ``in'').  For example, an object $o$ satisfies $\tuple{{\rm dept.id}, {\rm in}, \{{\rm CompSci}\}}$ if the value obtained starting from $o$ and following (dereferencing) the dept field and then the id field equals CompSci.
In examples, conditions are usually written using mathematical notation as syntactic sugar, with ``$\in$'' for ``in'' and ``$\ni$'' for ``contains''.  For example, $\tuple{{\rm dept.id}, {\rm in}, \{{\rm CompSci}\}}$ is more nicely written as ${\rm dept} \in \{{\rm CompSci}\}$. Note that the path is simplified by omitting the ``id'' field since all non-Boolean paths end with ``id'' field.  Also, ``='' is used as syntactic sugar for ``in'' when the constant is a singleton set; thus, the previous example may be written as dept=CompSci.

A {\em constraint} is a set, interpreted as a conjunction, of atomic constraints or their negations.  Informally, an atomic constraint expresses a relationship between the requesting subject and the requested resource, by relating the values of paths starting from each of them.  An {\em atomic constraint} is a tuple $\tuple{p_1, \op, p_2}$, where $p_1$ and $p_2$ are paths (possibly the empty sequence), and $\op$ is one of the following five operators: equal, in, contains, supseteq, subseteq.  
Implicitly, the first path is relative to the requesting subject, and the second path is relative to the requested resource.  The empty path represents the subject or resource itself.  For example, a subject $s$ and resource $r$ satisfy $\tuple{{\rm specialties}, {\rm contains}, {\rm topic}}$ if the set $s$.specialties contains the value $r$.topic.
In examples, constraints are written using mathematical notation as syntactic sugar, with ``$=$'' for ``equal'',``$\supseteq$'' for ``supseteq'', and ``$\subseteq$'' for ``subseteq''.

A {\em rule} is a tuple $\langle${\it subjectType}, {\it subjectCondition}, {\it resourceType}, {\it resourceCondition}, {\it constraint}, {\it actions}$\rangle$, where {\it subjectType} and {\it resourceType} are class names, {\it subjectCondition} and {\it resourceCondition} are conditions, {\it constraint} is a constraint, {\it actions} is a set of actions.  
A rule must satisfy several well-formedness requirements \cite{bui19mining}.
For a rule $\rho=\tuple{st, sc, rt, rc, c, A}$, let 
$\scond(\rho)=sc$, 
$\rcond(\rho)=rc$, $\con(\rho)=c$, and $\acts(\rho)=A$.

In the example rules, we prefix paths in conditions and constraints that start from the subject and resource with ``subject'' and ``resource'', respectively, to improve readability. 
For example, the e-document case study \cite{bui19mining,decat14edocShort} involves a  bank whose policy contains the rule: A project member can read all sent documents regarding the project.  Using syntactic sugar, this is written as
$\langle\,$Employee, subject.employer = LargeBank, Document, true, subject.workOn.relatedDoc $\ni$ resource, \{read\}$\rangle$, where Employee.workOn is the set of projects the employee is working on, and Project.relatedDoc is the set of sent documents related to the project.



The {\em type of a path} $p$ 
is the type of the last field in the path.  The {\em multiplicity of a path} $p$ 
is ``one'' if all fields on the path have multiplicity one, is many if any field on the path has multiplicity many, and is optional otherwise.  Given a class model, object model, object $o$, and path $p$, let $\nav(o,p)$ be the result of navigating (a.k.a. following or dereferencing) path $p$ starting from object $o$. If the navigation encounters {\tt unknown}, the result is {\tt unknown} if $p$ has multiplicity one or optional, and is a set of values containing {\tt unknown} (and possibly other values) if $p$ has multiplicity many. Otherwise, the result might be {\tt None}, an atomic value, or (if $p$ has multiplicity many) a set of values.  Aside from the extension to handle unknown, this is like the semantics of path navigation in UML's Object Constraint Language\footnote{\url{http://www.omg.org/spec/OCL/}}.

The truth value of an atomic condition $ac=\tuple{p, \op, \val}$ for an object $o$, denoted $\eval(o,ac)$, is defined as follows. If $p$ has multiplicity one (or optional) and $\nav(o,p)$ is {\tt unknown}, then $\eval(o,ac)=U$. If $p$ has multiplicity one (or optional) and $\nav(o,p)$ is known, then $\eval(o,ac)=T$ if $\nav(o,p) \in \val$, and $\eval(o,ac)=F$ otherwise. If $p$ has multiplicity many, then $\eval(o,ac)=T$ if $\nav(o,p) \ni \val$; otherwise, $\eval(o,ac)=F$ if $\nav(o,p)$ does not contain {\tt unknown}, and $\eval(o,ac)=U$ if it does.  Note that the operator $\op$ is not used explicitly in this definition, because $\op$ is uniquely determined by the multiplicity of $p$.  Next, we define $\eval$ for negated atomic conditions.  If $\eval(o,ac)=T$ and $\nav(o,p)$ is a set containing {\tt unknown}, then $\eval(o, \neg ac)=U$; otherwise, $\eval(o, \neg ac) = \neg \eval(o, ac)$, where $\neg$ denotes negation in three-valued logic \cite{3vlogic}. 

The truth value of an atomic constraint $ac=\tuple{p_1, \op, p_2}$ for a pair of objects $o_1, o_2$, denoted $\eval(o_1, o_2, ac)$, is defined as follows.  If no {\tt unknown} value is encountered during navigation, then $\eval(o_1, o_2, ac)=T$ if $(\op={\rm equal} \land \nav(o_1,p_1) = \nav(o_2,p_2)) \lor (\op={\rm in} \land \nav(o_1,p_1) \in \nav(o_2,p_2)) \lor (\op={\rm contains} \land \nav(o_1,p_1) \ni \nav(o_2,p_2)) \lor (\op={\rm supseteq} \land \nav(o_1,p_1) \supseteq \nav(o_2,p_2)) \lor (\op={\rm subseteq} \land \nav(o_1,p_1) \subseteq \nav(o_2,p_2))$, otherwise $\eval(o_1, o_2, ac)=F$. If $\nav(o_1,p_1)$ and $\nav(o_2,p_2)$ both equal {\tt unknown}, then $\eval(o_1, o_2, ac)=U$. If either of them is {\tt unknown} and $\op \in \{{\rm equal}, {\rm subseteq}, {\rm supseteq}\}$, then $\eval(o_1, o_2, ac)=U$. If either of them is {\tt unknown} and $\op \in \{{\rm in}, {\rm contains}\}$ (hence the other one is a set possibly containing {\tt unknown}), the truth value is defined similarly as in the corresponding case for atomic conditions. The truth value of negated atomic constraints is defined similarly as for negated atomic conditions. 

We extend $\eval$ from atomic conditions to conditions using conjunction (in three-valued logic): $\eval(o, \{ac_1,\ldots,ac_n\})=\eval(o,ac_1) \land \cdots \land \eval(o,ac_n)$.  We extend $\eval$ from atomic constraints to constraints in the same way.  An object or pair of objects {\em satisfies} a condition or constraint if $c$ has truth value $T$ for it.


An {\em SRA-tuple} is a tuple $\tuple{s, r, a}$, where the subject $s$ and resource $r$ are objects, and $a$ is an action, representing (depending on the context) authorization for $s$ to perform $a$ on $r$ or a request to perform that access.
An SRA-tuple $\tuple{s, r, a}$ {\em satisfies} a rule $\rho=\langle st, sc, rt,$ $rc, c, A\rangle$
if
$\type(s)=st \land \eval(s,sc)=T \land \type(r)=rt  \land \eval(r,rc)=T  \land \eval(\tuple{s,r},c)=T \land a \in A$.  The {\em meaning} of a rule $\rho$, denoted $\mean{\rho}$, is the set of SRA-tuples that satisfy it.
The {\em meaning} of a ReBAC policy $\pi$, denoted $\mean{\pi}$, is the union of the meanings of its rules.


\section{The Problem: ReBAC Policy Mining with Unknowns}
\label{sec:problem}

We adopt Bui et al.'s definition of the ReBAC policy mining problem and extend it to include unknown attribute values.  The ABAC policy mining problem is the same except it requires the mined policy to contain paths of length at most 1. 

An {\em access control list (ACL) policy} is a tuple $\tuple{\cm, \om, \Act, \spa}$, where $\cm$ is a class model, $\om$ is an object model that might contains unknown attribute values, $\Act$ is a set of actions, and $\spa\subseteq \om\times \om \times \Act$ is a set of SRA tuples representing authorizations. Conceptually, $\spa$ is the union of ACLs.  An ReBAC policy $\pi$ is {\em consistent} with an ACL policy $\langle\cm, \om,$ $\Act,$ $\spa\rangle$ if they have the same class model, object model, actions, and $\mean{\pi} = \spa$.

Among the ReBAC policies consistent with a given ACL policy $\pi_0$, the most desirable ones are those that satisfy the following two criteria.  (1) The ``id'' field should be used only when necessary, i.e., only when every ReBAC policy consistent with $\pi_0$ uses it, because uses of it make policies identity-based and less general.  (2) The policy should have the best quality as measured by a given policy quality metric $\Qpol$, expressed as a function from ReBAC policies to natural numbers, with small numbers indicating high quality.

The {\em ReBAC policy mining problem with unknown attribute values} is: given an ACL policy $\pi_0=\langle \cm, \om,$ $\Act, \spa\rangle$, where the object model $\om$ might contain unknown attribute values, and a policy quality metric $\Qpol$, find a set $\Rules$ of rules such that the ReBAC policy $\pi=\tuple{\cm, \om, \Act, \Rules}$ is consistent with $\pi_0$, uses the ``id'' field only when necessary, and has the best quality, according to $\Qpol$, among such policies.

The policy quality metric that our algorithm aims to optimize is {\em weighted structural complexity} (WSC), a generalization of policy size \cite{bui19mining}.  
WSC is a weighted sum of the numbers of primitive elements of various kinds that appear in a rule or policy.  It is defined bottom-up.   The $\wsc$ of an atomic condition $\tuple{p, \op, \val}$ is $|p| + |\val|$, where $|p|$ is the length of path $p$, and $|\val|$ is 1 if $\val$ is an atomic value and is the cardinality of $\val$ if $\val$ is a set.  The $\wsc$ of an atomic constraint $\tuple{p_1, \op, p_2}$ is $|p_1|+|p_2|$. The $\wsc$ of a negated atomic condition or constraint $c$ is 1 + $\wsc(c)$.  The WSC of a rule $\rho$, denoted $\wsc(\rho)$, is the sum of the WSCs of the atomic conditions and atomic constraints in it, plus the cardinality of the action set (more generally, it is a weighted sum of those numbers, but we take all of the weights to be 1). The WSC of a ReBAC policy $\pi$, denoted $\wsc(\pi)$, is the sum of the $\wsc$ of its rules. 


\section{ReBAC Policy Mining Algorithm}
\label{sec:mining-algorithm}

This section presents our ReBAC policy mining algorithms, DTRMU$^-$ and DTRMU.  
They have two main phases. The first phase learns a decision tree that classifies authorization requests as permitted or denied, and then constructs a set of candidate rules from the decision tree.  The second phase improves the policy by merging and simplifying the candidate rules and optionally removing negative atomic conditions/constraints from them.

\subsection{Phase~1:~Learn~Decision~Tree~and Extract Rules}
\label{sec:algorithm:dt}


\subsubsection{Problem decomposition.} 

We decompose the problem based on the subject type, resource type, and action.  Specifically, for each type $C_s$, type $C_r$, and action $a$ such that $\spa$ contains some SRA tuple with a subject of type $C_s$, a resource of type $C_r$, and action $a$, we learn a separate DNF formula $\phi_{C_s, C_r, a}$ to classify SRA tuples with subject type $C_s$, resource type $C_r$, and action $a$.  The decomposition by type is justified by the fact that all SRA tuples authorized by a rule contain subjects with the same subject type and resources with the same resource type.  Regarding the decomposition by action, the first phase of our algorithm generates rules that each contain a single action, but the second phase merges similar rules and can produce rules that authorize multiple actions.

\subsubsection{Construct labeled feature vectors.} 

To apply our formula-learning algorithm, we first need to extract sets of features and feature vectors from an input ACL policy. We use the same approach as described in \cite{bui20decision}. 

A {\em feature} is an atomic condition (on the subject or resource) or atomic constraint satisfying user-specified limits on lengths of paths in conditions and constraints.  We define a mapping from feature vectors to three-valued logic labels: given an SRA tuple $\tuple{s,r,a}$, we create a feature vector (i.e., a vector of the three-valued logic truth values of features evaluated for subject $s$ and resource $r$) and map it to $T$ if the SRA tuple is permitted (i.e., is in $\spa$) and to $F$ otherwise.  We do not label any feature vector with $U$, since the set of authorizations $\spa$ in the input ACL policy is assumed to be complete, according to the problem definition in Section \ref{sec:problem}.

\begin{table*}[htb]
\resizebox{\textwidth}{!}{%
\begin{tabular}{|l|l|l|c|c|c|c|l|}
\hline
\multicolumn{1}{|c|}{\multirow{2}{*}{FV\_id}} &
  \multicolumn{1}{c|}{\multirow{2}{*}{sub\_id}} &
  \multicolumn{1}{c|}{\multirow{2}{*}{res\_id}} &
  \multicolumn{4}{c|}{Features} &
  \multicolumn{1}{c|}{\multirow{2}{*}{Label}} \\ \cline{4-7}
\multicolumn{1}{|c|}{} &
  \multicolumn{1}{c|}{} &
  \multicolumn{1}{c|}{} &
  sub.dept   = res.dept &
  sub.dept   = CS &
  res.dept   = CS &
  res.type   = Handbook &
  \multicolumn{1}{c|}{} \\ \hline
1 & CS-student-1 & CS-doc-1 & $U$ & $T$ & $U$ & $T$ & $T$\\ \hline
2 & CS-student-1 & CS-doc-2 & $T$ & $T$ & $T$ & $U$ & $T$\\ \hline
3 & CS-student-1 & CS-doc-3 & $U$ & $T$ & $U$ & $U$ & $F$\\ \hline
4 & EE-student-1 & CS-doc-1 & $U$ & $U$ & $U$ & $T$ & $T$\\ \hline
5 & EE-student-1 & CS-doc-2 & $U$ & $U$ & $T$ & $U$ & $F$\\ \hline
6 & EE-student-1 & CS-doc-3 & $U$ & $U$ & $U$ & $U$ & $F$\\ \hline
\end{tabular}%
}\smallskip
\caption{Extracted features and feature vectors for the sample policy. FV\_id is a unique ID assigned to the feature vector. sub\_id and res\_id are the subject ID and resource ID, respectively.  Features that are conditions on sub\_id or res\_id are not shown.  The labels specify whether a student has permission to read a document ($T$=permit, $F$=deny). 
}
\label{tab:sample-policy}
\end{table*}

Table \ref{tab:sample-policy} shows a set of labeled feature vectors for our running example, which is a ReBAC policy containing two student objects, with IDs CS-student-1 and EE-student-1, and three document objects, with IDs CS-doc-1, CS-doc-2 and CS-doc-3. Each student object has a field ``dept'' specifying the student's department. Each document object has a field ``dept'' specifying which department it belongs to, and a field ``type'' specifying the document type. The field values are CS-student-1.dept = {\tt CS}, EE-student-1.dept = {\tt unknown}, CS-doc-1.dept = CS-doc-3.dept = {\tt unknown}, CS-doc-2.dept = {\tt CS}, CS-doc-1.type = {\tt Handbook}, and CS-doc-2.type = CS-doc-3.type = {\tt unknown}.  The labels are consistent with the ReBAC policy containing these two rules: (1) A student can read a document if the document belongs to the same department as the student, and (2) every student can read handbook documents. Formally, the rules are (1) $\langle\,$Student, true, Document, true, subject.dept $=$ resource.dept, \{read\}$\rangle$, and (2) $\langle\,$Student, true, Document, resource.type = Handbook, true, \{read\}$\rangle$. Note that this is also an ABAC policy, since all paths have length 1. 

The feature vectors constructed to learn $\phi_{C_s, C_r, a}$ include only features appropriate for subject type $C_s$ and resource type $C_r$, e.g., the path in the subject condition starts with a field in class $C_s$. The set of labeled feature vectors used to learn $\phi_{C_s, C_r, a}$ contains one feature vector generated from each possible combination of a subject of type $C_s$ (in the given object model) and a resource of type $C_r$.  We also use the optimizations described in \cite[Section 5.1]{bui20decision} to discard some ``useless'' features, namely, features that have the same value in all feature vectors, and sets of features equivalent to simpler sets of features.  For the running example, Table \ref{tab:sample-policy} shows the feature vectors for $C_s = {\rm Student}, C_r = {\rm Document}, a = {\rm read}$.

\subsubsection{Learn a formula.}
\label{sec:mining-apply-general-algorithm}

After generating the labeled feature vectors, we apply the formula-learning algorithm in Section \ref{sec:mining-algorithm:extract-formula}.  We do not explicitly check the monotonicity of the set of labeled feature vectors.  Instead, after constructing each formula, we directly check whether it is valid (it will always cover the given set of labeled feature vectors); this is necessary because, if the set of labeled feature vectors is not monotonic, disjuncts added by the loop over {\it uncov} might be invalid.  This approach has two benefits: it is computationally cheaper because it requires iterating over feature vectors (or, equivalently, subject-resource-action tuples) individually, whereas monotonicity requires considering pairs of feature vectors; and it provides an end-to-end correctness check as well as an implicit monotonicity check.

To help the formula-learning algorithm produce formulas that lead to rules with lower WSC, we specialize the scoring metric used to choose a feature to test at each node.  Specifically, we use information gain as the primary metric, but we extend the metric to use WSC (recall that WSC of atomic conditions and atomic constraints is defined in Section \ref{sec:problem}) as a tie-breaker for features that provide the same information gain.


Specialized treatment of conditions on the ``id'' attribute, e.g., ${\rm subject.id} = \mbox{CS-student-1}$ is also beneficial.  Recall from Section \ref{sec:problem} that such conditions should be used only when needed.  Also, we expect that they are rarely needed.  We consider two approaches to handling them.   In the first approach, we first run the formula-learning algorithm on feature vectors that do not contain entries for these conditions; this ensures those conditions are not used unnecessarily, and it can significantly reduce the running time, since there are many such conditions for large object models.  That set of feature vectors is not necessarily monotonic, so the learned formula might not be valid; this will be detected by the validity check mentioned above.   If it is not valid, we generate new feature vectors that include these conditions and run the formula-learning algorithm on them.

In the second approach, we run a modified version of the formula-learning algorithm on feature vectors that do not contain entries for these conditions.  The modification is to the loop over {\it uncov}: for each feature vector $v$ in {\it uncov}, it adds the conjunction ${\rm subject.id}=id_s \land {\rm resource.id}=id_r$ to $D$, where $s$ and $r$ are the subject and resource, respectively, for which $v$ was generated, and $id_s$ and $id_r$ are their respective IDs.  The disadvantage of this approach is that it can sometimes use conditions on {\it id} when they are not strictly needed; the advantage of this approach is that it can sometimes  produce policies with smaller WSC, because the modified version of the loop over {\it uncov} produces conjunctions with few conjuncts, while the original version of the loop over {\it uncov} produces conjunctions with many conjuncts (though some conjuncts may be removed by simplifications in phase 2).

In practice, both approaches usually produce the same result, because, even when conditions on ''id'' are omitted, the first top-level loop in the formula-learning algorithm usually succeeds in covering all feature vectors.

\begin{figure}[htb]
  \centering
\includegraphics[width=0.5\textwidth]{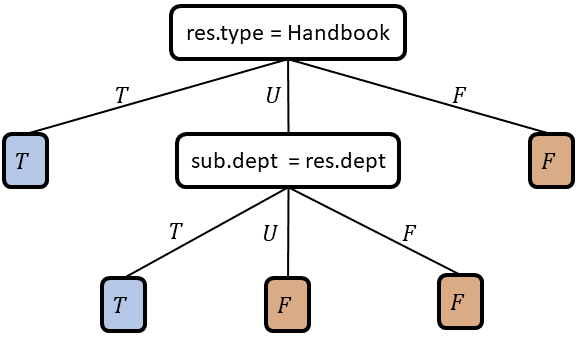}
  \caption{Multi-way decision tree for the running example.}
  \label{fig:sample-dtrmu}
\end{figure}

Figure \ref{fig:sample-dtrmu} shows the learned multi-way decision tree for the set of labeled feature vectors in Table \ref{tab:sample-policy}. Internal nodes and leaf nodes are represented by unfilled and filled boxes, respectively. The conjunctions conj($p$) generated from paths from the root to a leaf labeled $T$ are (1) ${\rm res.type = Handbook}$ and (2) $\langle{\rm res.type = Handbook\rangle} = U \land {\rm sub.dept = res.dept}$.  Note that, for convenience, a formula containing a single condition is considered to be a (degenerate) kind of conjunction.

The algorithm tries to eliminate the condition $\langle{\rm res.type = Handbook}\rangle = U$ in conjunction (2).  Removing that condition leaves the (one-element) conjunction ${\rm sub.dept = res.dept}$, which is still valid WRT to the set of feature vectors in Table \ref{tab:sample-policy}, so the algorithm replaces conjunction (2) with ${\rm sub.dept = res.dept}$ in $D$.  The first top-level loop in the algorithm succeeds in covering all feature vectors in $S$.  Thus, the learned formula $\phi_{{\rm Student, Document, read}}$ is $({\rm res.type = Handbook}) \lor ({\rm sub.dept = res.dept})$.

\subsubsection{Extract rules.}
\label{sec:mining-algorithm:extract-rules}

We convert the formula into an equivalent set of rules and add them to the candidate mined policy. For each conjunction $c$ in the formula $\phi_{C_s, C_r, a}$, we create a rule with subject type $C_s$, resource type $C_s$, action $a$, and with $c$'s conjuncts as atomic conditions and atomic constraints.
For the running example, the formula $\phi_{{\rm Student, Document, read}}$ has two (degenerate) conjunctions, and the algorithm successfully extracts the two desired rules given above in the description of Table \ref{tab:sample-policy}.


\subsection{Phase 2: Improve the Rules}
\label{sec:algorithm:improve}

Phase 2 has two main steps: eliminate negative features, and merge and simplify rules. We adopt these steps from DTRM. We give brief overviews of these steps in this paper, and refer the reader to \cite{bui20decision} for additional details. 

\subsubsection{Eliminate Negative Features.}
\label{sec:algorithm:elim-neg}

This step is included only in DTRMU, in order to mine rules without negation.  This step is omitted from DTRMU$^-$.  It eliminates each negative feature in a rule $\rho$ by removing the negative feature (if the resulting rule is valid) or replacing it with one or more positive feature(s).

\subsubsection{Merge and Simplify Rules.}
\label{sec:algorithm:simplify}

This step attempts to merge and simplify rules using the same techniques as \cite{bui20decision} (e.g., removing atomic conditions and atomic constraints when this preserves validity of the rule, eliminating overlap between rules, and replacing constraints with conditions), extended with one additional simplification technique: If an atomic condition on a Boolean-valued path $p$ has the form $p \ne F$ or $p \ne T$, it is replaced with $p = T$ or $p = F$, respectively.

\subsubsection{Naively applying DTRM.}
\label{sec:DTRM-examples}

One might wonder whether DTRM$^-$ can be used to mine policies, by assuming that features involving unknown attribute values evaluate to $F$ (instead of $U$). Although there is no reason to believe that this will work, it is easy to try, so we did.  For the running example, DTRM$^-$ produces two rules:$\langle\,$Student, true, Document, res.type = Handbook, true, \{read\}$\rangle$ and $\langle\,$Student, true, Document, res.type $\neq$ Handbook, sub.dept = res.dept, \{read\}$\rangle$.  This  policy is incorrect, because it does not cover feature vector 2 in Table \ref{tab:sample-policy}, i.e., it prevents CS-student-1 from reading CS-doc-2.

\section{Evaluation Methodology}
\label{sec:evaluation-methodology}

\begin{figure*}[htb]
\begin{tabular}[b]{|l|l|l|l|l|}
\hline
\multicolumn{1}{|c|}{Policy\_$N$} & \multicolumn{1}{c|}{\#obj} & \multicolumn{1}{c|}{\#field} & \multicolumn{1}{c|}{\#FV} & \multicolumn{1}{c|}{\#rule} \\ \hline
EMR\_15 & 353 & 877 & 4134 & 6 \\ \hline
healthcare\_5 & 736 & 1804 & 42121 & 8 \\ \hline
healthcare\_5$^-$ & 736 & 1875 & 42121 & 8 \\ \hline
project-mgmt\_5 & 179 & 296 & 4080 & 10 \\ \hline
project-mgmt\_10$^-$ & 376 & 814 & 23627 & 10 \\ \hline
university\_5 & 738 & 926 & 83761 & 10 \\ \hline
e-document\_75 & 284 & 1269 & 31378 & 39 \\ \hline
eWorkforce\_10 & 412 & 1124 & 14040 & 19 \\ \hline

\end{tabular}
\hspace*{0.5em}\includegraphics[width=0.45\textwidth]{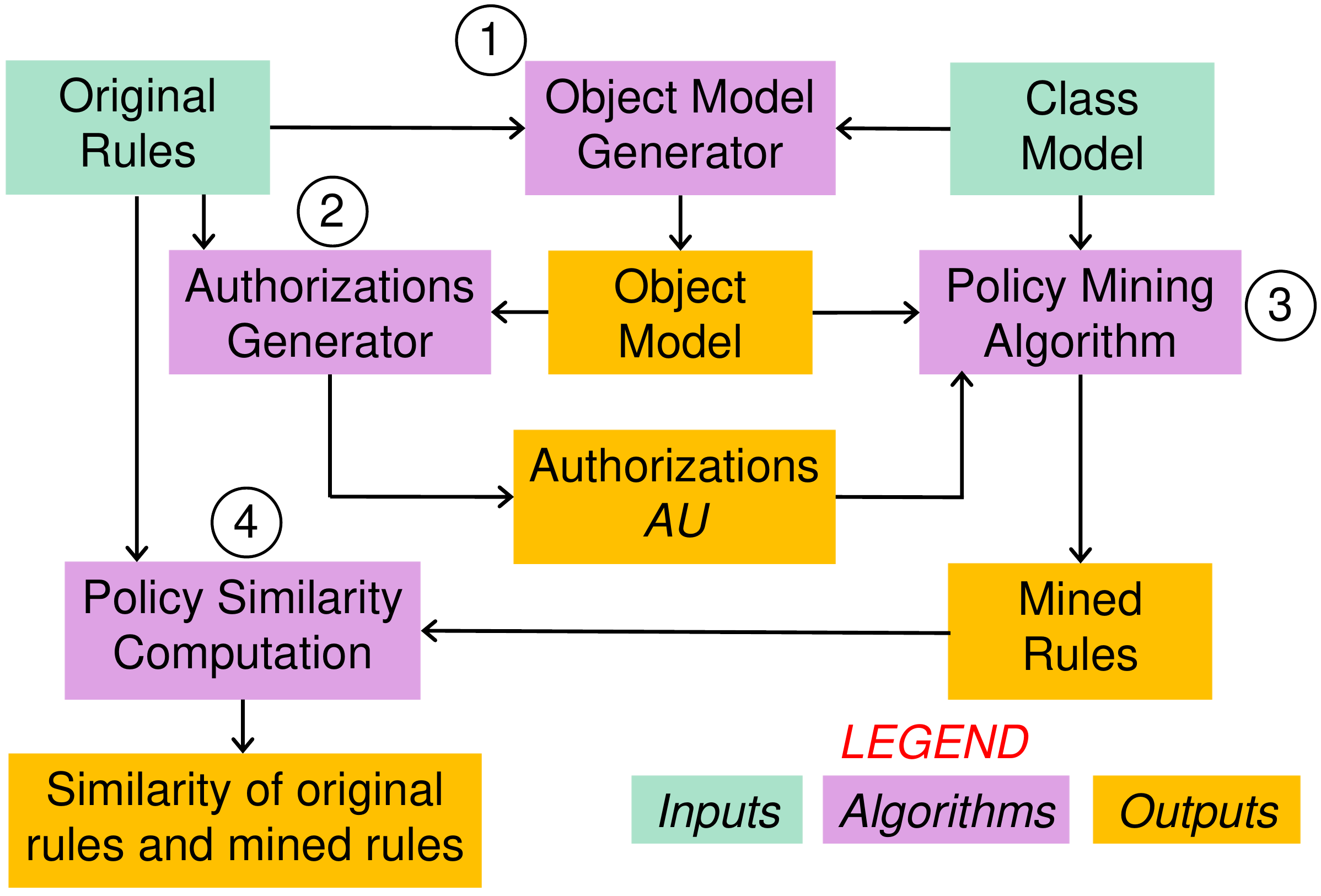}
\smallskip
\caption{Left: Policy sizes. For the given value of the object model size parameter $N$ (after the underscore in the policy name), \#obj is the average number of objects in the object model, and \#field is the average number of fields in the object model, i.e., the sum over objects $o$ of the number of fields in $o$.  \#FV is the number of feature vectors (i.e., labeled SRA tuples) that the algorithms generate to learn a  formula.  Averages are over 5 pseudorandom object models for each policy. ``healthcare\_5$^-$'' and ``project-mgmt\_10$^-$'' are the policies with negations that we generated. Right: Evaluation methodology; reproduced from \cite{bui20decision}.}
\label{tab:policy-size}
\label{fig:eval-method}
\end{figure*}

We adopt Bui et al.'s methodology for evaluating policy mining algorithms \cite{bui19sacmat}.  It is depicted in Figure \ref{fig:eval-method}.  It takes a class model and a set of ReBAC rules as inputs.  The methodology is to generate an object model based on the class model (independent of the ReBAC rules), compute the authorizations $\spa$ from the object model and the rules, run the policy mining algorithm with the class model, object model, and $\spa$ as inputs, and finally compare the mined policy rules with the simplified original (input) policy rules, obtained by applying the simplifications in Section \ref{sec:algorithm:simplify} to the given rules.  Comparison with the simplified original policy is a more robust measure of the algorithm's ability to discover high-level rules than comparison with the original policy, because the original policy is not always the simplest.  If the mined rules are similar to the simplified original rules, the policy mining algorithm succeeded in discovering the desired ReBAC rules that are implicit in $\spa$.

\subsection{Datasets}
\label{sec:datasets}

We use four sample policies developed by Bui et al. \cite{bui19mining}.  One is for electronic medical records (EMR), based on the EBAC policy in \cite{bogaerts15entityShort}, translated to ReBAC; the other three are for healthcare, project management, and university records, based on ABAC policies in \cite{xu15miningABAC}, generalized and made more realistic, taking advantage of ReBAC's expressiveness.  These  policies are non-trivial but relatively small. 

We also use Bui et al.'s translation into ORAL2$^-$ \cite{bui20decision} of two large case studies developed by Decat, Bogaerts, Lagaisse, and Joosen based on the access control requirements for Software-as-a-Service (SaaS) applications offered by real companies \cite{decat14edoc,decat14workforce}.  One is for a SaaS multi-tenant e-document processing application; the other is for a SaaS workforce management application provided by a company that handles the workflow planning and supply management for product or service appointments (e.g., install or repair jobs).

More detailed descriptions of these policies are available in \cite{bui20decision}.  The ABAC or ReBAC versions of these policies, or variants of them, have been used as benchmarks in several  papers on policy mining, including \shortonly{\cite{medvet2015,bui18mining,iyer2018,iyer2019,bui20decision}}\fullonly{\cite{medvet2015,bui18mining,iyer2018,karimi2018,iyer2019,bui20decision}}.



These sample policies and the case studies do not include any rules with negations.  Therefore, we created modified versions of the healthcare and project management policies that include some rules with negation; the names of the modified version end with ``$^-$''.   For the healthcare policy, we add a new attribute ``COIs'' in the Patient class to specify the physicians or nurses who have a conflict of interest with the patient, and in the rules that give any permission on a patient's record to a physician or nurse, we add the constraint ${\rm subject} \notin {\rm resource.patient.COIs}$.  For the project management policy, we add a new attribute ``status'' in the Task class with possible values not\_started,  in\_progress, and completed.  In the rules that give permission to change the cost, schedule, or status of a task, we add the condition $\rm{resource.status} \neq completed$. 

The object models are generated by policy-specific pseudorandom algorithms designed to produce realistic object models, by creating objects and selecting their attribute values using appropriate probability distributions.  These algorithms are parameterized by a size parameter $N$; for most classes, the number of instances is selected from a normal distribution whose mean is linear in $N$.  We use the same object model generators as Bui and Stoller \cite{bui20decision}, which are slightly modified versions of the object model generators described and used in \cite{bui19sacmat,bui19mining}, which are available online.\footnote{\url{https://www.cs.stonybrook.edu/~stoller/software/}}
Note that, in these object models, all attribute values are known.
The table in Figure \ref{tab:policy-size} shows several metrics of the size of the rules, class model, and object model in each policy.

\subsection{Policy Similarity Metrics}
\label{sec:policy-sim-metrics}

We evaluate the quality of the generated policy primarily by its {\em syntactic similarity} and {\em policy semantic similarity} to the simplified original policy.  These metrics are first defined in \cite{xu15miningABAC,bui18mining} and adapted in \cite{bui20decision} to take negation into account.   They are normalized to range from 0 (completely different) to 1 (identical).  They are based on Jaccard similarity of sets, defined by $J(S_1, S_2) = |S_1\intersect S_2| \,/\, |S_1 \union S_2|$.  For convenience, we extend $J$ to apply to single values: $J(v_1,v_2)$ is 1 if $v_1=v_2$ and 0 otherwise.

{\em Syntactic similarity} of policies measures the syntactic similarity of rules in the policies, based on the fractions of types, conditions, constraints, and actions that rules have in common. The {\em syntactic similarity of rules} is defined bottom-up as follows.    
For an atomic condition $ac$, let $\getsign(ac)$, $\getpath(ac)$, and $\getval(ac)$ denote its sign (positive or negative), its path, and its value (or set of values), respectively.  Syntactic similarity of atomic conditions $ac_1$ and $ac_2$, $\acSim(ac_1, ac_2)$, is 0 if they contain different paths, otherwise it is the average of
 $J(\getsign(ac_1), \getsign(ac_2))$, $J(\getpath(ac_1), \getpath(ac_2))$,
and $J(\getval(ac_1), \getval(ac_2)))$; we do not explicitly compare the operators, because atomic conditions with the same path must have the same operator, since the operator is uniquely determined by the multiplicity of the path.  For a set $S$ of atomic conditions, let $\paths(S)= \{ \getpath(ac)\;|\; ac \in S \}$.  For sets $S_1$ and $S_2$ of atomic conditions,
\begin{displaymath}
\synSim(S_1, S_2) = |\paths(S_1) \union \paths(S_2)|^{-1} \acmonly{\!}\sum_{ac_1 \in S_1, ac_2 \in S_2} \acmonly{\!\!}\acSim(ac_1, ac_2)
\end{displaymath}
The {\em syntactic similarity of rules} $\rho_1=\langle st_1, sc_1, rt_1, rc_1, c_1, A_1\rangle$ and $\rho_2=\tuple{st_2, sc_2,$ $rt_2, rc_2, c_2, A_2}$ is
$\synSim(\rho_1, \rho_2) = \average(%
    J(st_1, st_2), 
    \synSim(sc_1,$ $sc_2), 
    J(rt_1, rt_2),$ \linebreak[4]
    $\synSim(rc_1, rc_2), J(c_1, c_2),  J(A_1, A_2))$.
%
The {\em syntactic similarity of policies} $\pi_1$ and $\pi_2$, denoted $\synSim(\pi_1$, $\pi_2$), is the average, over rules $\rho$ in $\pi_1$, of the syntactic similarity between $\rho$ and the most similar rule in $\pi_2$. 

The {\em semantic similarity of polices} measures the fraction of authorizations that the policies have in common.  Specifically, the {\em semantic similarity} of policies $\pi_1$ and $\pi_2$ is $J(\mean{\pi_1},\mean{\pi_2})$. 


\section{Evaluation Results}
\label{sec:evaluation-results}

We performed two series of experiments.  The first series of experiments compares our algorithms with Bui and Stoller's DTRM and DTRM$^-$ algorithms (which are state-of-the-art, as discussed in Section \ref{sec:intro}), and shows that, {\em on policies where all attribute values are known, our algorithms are equally effective at discovering the desired ReBAC rules, produce policies with the same quality, and have comparable running time}.  The second series of experiments, on policies containing a varying percentage of unknown values, shows that {\em our algorithms are effective at discovering the desired ReBAC rules, even when a significant percentage of attribute values are unknown}.



We implemented our formula-learning algorithm in Python, on top of Esmer's implementation of the C4.5 decision-tree learning algorithm\footnote{\url{https://github.com/barisesmer/C4.5}}.
Bui and Stoller's implementation of DTRM$^-$ \cite{bui20decision} uses the optimized version of the CART decision-tree learning algorithm provided by the scikit-learn library\footnote{\url{https://scikit-learn.org/stable/modules/tree.html}}; we could not use it, because it supports only binary trees.  Since Esmer's implementation of C4.5 supports only information gain as the feature scoring metric, we chose it as the scoring metric in scikit-learn when running DTRM and DTRM$^-$, which originally used the default scoring metric, which is gini index.  This change had a negligible effect on the algorithms' output (no effect for all policies except e-document\_75, for which it improved the results slightly) and allows a fairer comparison of DTRM and DTRM$^-$ with DTRMU and DTRMU$^-$.  We re-used Bui and Stoller's implementation of Phase 2, with the small extension in Section \ref{sec:mining-algorithm:extract-rules}.  
When generating feature vectors, we use the same path length limits ({\it cf.} Section \ref{sec:algorithm:dt}) as in \cite{bui19mining,bui19sacmat} for all algorithms. We set the value of the $max\_iter$ parameter in the formula-learning algorithm to 5. We ran DTRM$^-$ and DTRMU$^-$ on the policies containing rules with negation (healthcare\_5$^-$ and project-mgmt\_10$^-$), and we ran DTRM and DTRMU on the other policies.  All experiments were run on Windows 10 on an Intel i7-6770HQ CPU.


\subsection{Comparison with DTRM and DTRM$^-$}
\label{sec:exper:comparison}

We compared our algorithms with DTRM and DTRM$^-$ using the datasets described in Section \ref{sec:datasets}.  We ran experiments on five object models for each policy and averaged the results. The standard deviations (SD) are reasonable, indicating that averaging over five object models for each data point is sufficient to obtain meaningful results.

All of these algorithms always mine policies that grant the same authorizations as the input ACL policies and thus achieve perfect {\em semantic similarity} for all datasets.  Our algorithms achieve almost exactly the same  {\em syntactic similarity} as DTRM and DTRM$^-$ when comparing mined rules with simplified original rules, as explained in Section \ref{sec:evaluation-methodology}.  DTRM and DTRMU both achieve the same results for average syntactic similarity: 1.0 (SD = 0) for healthcare\_5, project-mgmt\_5, and university\_5; 0.99 (SD = 0.01) for EMR\_15; 0.98 (SD = 0.01) for eWorkforce\_10; and 0.92 (SD = 0.02) for e-document\_75.  DTRMU$^-$ and DTRM$-$ both achieve 1.0 (SD = 0) average syntactic similarity for healthcare\_5$^-$ and project-mgmt\_10$^-$. For all of the datasets, our algorithms and theirs mine policies with the same average WSC.

The running times of DTRM and DTRM$^-$ are somewhat faster than our algorithms. Averaged over all policies, DTRM is 1.53 (SD = 0.20) times faster than DTRMU, and DTRM$^-$ is 1.72 (SD = 0.05) times faster than DTRMU$^-$. The difference in the running time comes mostly from the decision-tree learning step.  When there are no unknown values, our algorithms and DTRM and DTRM$^-$ are essentially the same at the algorithm level, aside from our algorithms having a very small overhead to check for unknowns.  Therefore, we attribute the difference in running time primarily to the use of different tree-learning libraries---Esmer's straightforward implementation of C4.5 used by our algorithms vs.\ the optimized version of CART in scikit-learn used by DTRM and DTRM$^-$.  Furthermore, C4.5 and CART are similar algorithms and should construct the same binary trees when applied to boolean feature vectors labeled with booleans (they handle continuous data differently), so the difference in running time is mainly due to implementation-level differences.


\subsection{Experiments with unknown attribute values}
\label{sec:exper:unknown}

We generated datasets with unknown attribute values by changing the values of pseudorandomly chosen fields to {\tt unknown} in the datasets used for the experiments in Section \ref{sec:exper:comparison}.  We introduce a scaling factor $s$ to vary how many {\tt unknown} values are introduced.  In each policy, for most fields $f$ of each class $C$, we pseudorandomly choose a probability $p$ in the range $[0.02 s, 0.05 s]$, and then, for each instance $o$ of $C$, we change the value of $f$ to {\tt unknown} with probability $p$.  This is done for all fields except a few manually classified as {\em required} or {\em important}.  For a required field, we take $p=0$, i.e., no instances are changed to unknown.  For an important field (i.e., one whose value is more likely to be known), we take $p = 0.01 s$.  For example, the university policy has one required field, Transcript.student (the student whose transcript it is), and one important field, Faculty.department.  For all policies, the number of required or important fields is less than 15\% of the total number of fields in the class model.

We ran experiments with $s=0$ (i.e., all attribute values are known, same datasets as in Section \ref{sec:exper:comparison}), 1, 2, and 3.  Averaged over all policies, the percentages of field values in the object model that are changed to {\tt unknown} are 3\%, 6\%, and 8\% for $s = 1$, 2, and 3, respectively.  Experimental results appear in Table \ref{tab:dtrmu-results} and are discussed below.


\begin{table*}[htb]
\resizebox{\textwidth}{!}{%
\begin{tabular}{|l|c|c|c|c|c|c|c|c|}
\hline
\multicolumn{1}{|c|}{\multirow{2}{*}{Policy}} &
  \multicolumn{2}{c|}{$s = 0$} &
  \multicolumn{2}{c|}{$s = 1$} &
  \multicolumn{2}{c|}{$s = 2$} &
  \multicolumn{2}{c|}{$s = 3$} \\ \cline{2-9} 
\multicolumn{1}{|c|}{} &
  Syn. Sim &
  Run Time &
  Syn. Sim &
  Slowdown &
  Syn. Sim &
  Slowdown &
  Syn. Sim &
  Slowdown \\ \hline
EMR\_15              & 0.99 & 76.19  & 0.99 & 2.28  & 1.00 & 9.52  & 0.99 & 6.46 \\ \hline
healthcare\_5        & 1.00 & 129.50 & 1.00 & 1.24  & 1.00 & 1.25  & 0.99 & 1.23 \\ \hline
project-mgmt\_5      & 1.00 & 3.81   & 1.00 & 1.08  & 1.00 & 1.21  & 1.00 & 1.45 \\ \hline
univeristy\_5        & 1.00 & 266.29 & 1.00 & 1.07  & 1.00 & 1.03  & 1.00 & 1.07 \\ \hline
eWorkforce\_10       & 0.98 & 96.40  & 0.96 & 11.88 & 0.93 & 8.41  & 0.94 & 9.13 \\ \hline
e-document\_75       & 0.92 & 420.27 & 0.93 & 8.03  & 0.93 & 15.77 & 0.91 & 9.41 \\ \hline
healthcare\_5$^-$    & 1.00 & 171.95 & 1.00 & 1.38  & 0.99 & 1.38  & 0.99 & 1.34 \\ \hline
project-mgmt\_10$^-$ & 1.00 & 38.04  & 1.00 & 1.39  & 0.99 & 1.71  & 0.99 & 1.74 \\ \hline
\end{tabular}%
}\smallskip
\caption{Experimental results for our algorithms on datasets with different values of scaling factor $s$. ``Syn. Sim'' is the average syntactic similarity achieved on each policy. ``Run time'' is measured in seconds.  For $s>0$, we report the slowdown relative to $s=0$, i.e., the ratio of the running time to the running time on the same policy with $s=0$.
}
\label{tab:dtrmu-results}
\end{table*}

\subsubsection{Policy Similarity and WSC.}

Our algorithms always mine policies that grant exactly the same authorizations as the input ACL policies and thus achieve perfect {\em semantic similarity} for all datasets.

For the sample policies (including the variants with negation), our algorithms achieve 0.99 or better average syntactic similarity for all four values of $s$.  For the case studies, DTRMU achieves 0.96, 0.93, and 0.94 syntactic similarity for eWorkforce\_10 with $s = 1$, 2, and 3, respectively; for e-document\_75, the results are 0.93, 0.93, and 0.91, respectively.  The standard deviations are less than 0.02 for all results, except for eWorkforce\_10 with $s=2$, where $SD=0.04$.  In short, we see that {\em unknown attribute values cause a small decrease in policy quality, but policy quality remains high} even with up to 8\% of field values set to {\tt unknown} (with $s=3$), and trend downward slowly as the percentage of unknowns increases.


Our algorithms generate policies with {\em the same or better (smaller) average WSC than the simplified input policies} for all datasets except e-document\_75, for which the average WSC of the mined policy is 20\%, 14\%, and 15\%  higher in experiments with $s = 1$, 2, and 3, respectively.  The standard deviations (over the 5 object models) for each policy are between 5\% and 9\% of the averages for EMR\_15, eWorkforce\_10, and e-document\_75; for other policies, the standard deviations are 0.

\subsubsection{Running Time.} 

Table \ref{tab:dtrmu-results} reports our algorithms' running times for $s=0$ and the slowdown (relative to $s=0$) for larger values of $s$.  This slowdown reflects the additional processing needed to handle unknown values.  Averaged over all policies, the average slowdown is 3.5, 5.0, and 4.0 for  $s = 1$, 2, and 3, respectively.  The median slowdown is 1.4, 1.5, and 1.6 for $s = 1$, 2, and 3, respectively.  

Our algorithms spend most of the time in phase 1, to learn decision trees and extract rules. The slowdown on a few policies is notably larger than the others, and the standard deviations in running time for those policies are also high, indicating that, for each of those policies, the algorithms take much longer on a few object models than on the others.  The larger slowdown for these object models is caused by additional time spent eliminating features involving the unknown.  In particular, several of the features involving unknown cannot be eliminated by {\it max\_iter} iterations of the top-level $\whileloop$ loop, so the $\forloop$ loop over {\it uncov} is executed to eliminate them; we use the second approach in Section \ref{sec:mining-apply-general-algorithm}, generating rules that use ``id''.
 On the positive side, these low-quality rules are removed in phase 2, and the algorithms still succeed in mining high-quality policies.


\section{Related Work}
\label{sec:related}


We discuss related work on policy mining.  As mentioned in Section \ref{sec:intro}, the primary distinction of our work is that {\em no related work on ReBAC or ABAC policy mining considers unknown attribute values}.  We are not aware of related work on learning concise formulas in three-valued logic.

\subsubsection{Related work on ReBAC policy mining.}

Bui et al. developed several ReBAC policy mining algorithms \cite{bui17mining,bui18mining,bui19mining,bui19sacmat,bui20decision}, the most recent and best of which are DTRM and DTRM$^-$ \cite{bui20decision}.  Our algorithms modify them to handle unknown attribute values.  Bui et al.'s algorithms in \cite{bui18mining} can mine ReBAC policies from incomplete and noisy information about permissions \cite{bui18mining}.  

Iyer et al. present algorithms, based on ideas from rule mining and frequent graph-based pattern mining, for mining ReBAC policies and graph transition policies \cite{iyer2019}.  Their policy mining algorithm targets a policy language that is less expressive than ORAL2$^-$, because it lacks set comparison operators and negation; furthermore, unlike ORAL2, it does not directly support Boolean attributes, and encoding them may be inefficient \cite{bui20decision}. 
Also, in Bui and Stoller's experiments, DTRM is faster and more effective than their algorithm \cite{bui20decision}.

Iyer et al. \cite{iyer2020active} present an algorithm for active learning of ReBAC policies from a black-box access control decision engine, using authorization queries and equivalence queries.  The algorithm is assumed to have access to complete information about attributes and relationships.

\subsubsection{Related work on ABAC Policy mining.}

Xu et al. proposed the first algorithm for ABAC policy mining \cite{xu15miningABAC} and a variant of it for mining ABAC policies from logs \cite{xu14miningABAClogs}.  Medvet et al. developed the first evolutionary algorithm for ABAC policy mining \cite{medvet2015}.  Iyer et al. developed the first ABAC policy mining algorithm that can mine ABAC policies containing deny rules as well as permit rules \cite{iyer2018}.\fullonly{  Karimi et al. proposed an ABAC policy mining algorithm that uses unsupervised learning based on $k$-modes clustering \cite{karimi2018}.}  Cotrini et al. proposed a new formulation of the problem of ABAC mining from logs and an algorithm based on APRIORI-SD, a machine-learning algorithm for subgroup discovery, to solve it \cite{sparselogs2018}.  Cotrini et al. also developed a ``universal'' access control policy mining algorithm framework, which can be specialized to produce policy mining algorithms for a wide variety of policy languages \cite{cotrini2019}; the downside, based on their experiments, is that the resulting algorithms achieve lower policy quality than customized algorithms for specific policy languages.  Law et al. present a scalable inductive logic programming algorithm and evaluate it for learning ABAC rules from logs \cite{law20fastlas}.




\articleonly{\paragraph{Acknowledgements.} \thanksText }


\svonly{
\begin{acknowledgements}
\thanksText
\end{acknowledgements}}


%
\ieeeonly{\bibliographystyle{IEEEtran}}\acmonly{\bibliographystyle{ACM-Reference-Format}}\lncsonly{\bibliographystyle{splncs04}}\articleonly{\bibliographystyle{alpha}}\svonly{\bibliographystyle{plain}}
\bibliography{references}

 \end{document}
